# Autoionization of water: does it really occur?

V. G. Artemov[1], A. A. Volkov[2], N. N. Sysoev[2], A. A. Volkov[1]

[1] A.M. Prokhorov General Physics Institute, Russian Academy of Sciences, 119991 Moscow, Russia

[2] Physical Department, M.V. Lomonosov Moscow State University, 119991, Moscow, Russia

E-mail: vartemov@bk.ru

**Abstract.** The ionization constant of water $K_w$ is currently determined on the proton conductivity $\sigma_0$ which is measured at frequencies lower than $10^7$ Hz. Here, we develop the idea that the high frequency conductivity $\sigma_\infty$ (~$10^{11}$ Hz), rather than $\sigma_0$ represents a net proton dynamics in water, to evaluate the actual concentration $c$ of $H_3O^+$ and $OH^-$ ions from $\sigma_\infty$. We find $c$ to be not dependent on temperature to conclude that i) water electrodynamics is due to a proton exchange between $H_3O^+$ (or $OH^-$) ions and neutral $H_2O$ molecules rather than spontaneous ionization of $H_2O$ molecules, ii) the common Kw (or pH) reflects the thermoactivation of the $H_3O^+$ and $OH^-$ ions from the potential of their interaction, iii) the lifetime of a target water molecule does not exceed parts of nanosecond.

PACS 61.20.Gy – Theory and models of liquid structure
PACS 77.22.-d – Dielectric properties of solids and liquids

## 1. Introduction

Proton conductivity $\sigma_0$ is a fundamental property of liquid water. It is commonly associated with the presence of $H_3O^+$ and $OH^-$ ions in water which are generated by autodissociation of $H_2O$ molecules [1-4]. Autodissociation is maintained by the reverse process of recombination. The phenomenon of dissociation-association of water molecules is a basis for the practically important hydrogen pH-concept [5].

The broken water molecule is typically restored in the femtoseconds time scale. With some probability, however, the separated ions are hydrated, their lifetime lasting to 70 ms [6, 7]. The lifetime of an $H_2O$ molecule is estimated to be of ~10 hours at 25 °C. This value results from the old spark experiments of Eigen [8].

This picture has been entrenched in physical chemistry. It feeds continuous MD simulations in which the dynamics of the hydrogen bond system and the mechanisms of the proton transfer are studied [6,7, 9-13]. Although a detailed understanding of separate microscopic mechanisms has been achieved, a satisfactory consistent theory, however, is still not developed [13].

Autoionzation of water is given quantitatively by a dissociation constant $K_w$ whose decimal logarithm is equal to 2pH. The problem of great practical importance is understanding of the $K_w$ non-monotonous temperature behavior – there is a minimum at around 220-240 °C. As is believed, the $K_w$ temperature dependence is due to hydration of the $H_3O^+$ and $OH^-$ ions [14, 15]. Generally, the description of the $K_w(T)$ is a complex multiparticle problem.

Experimentally, $K_w$ is obtained from the conductometric measurements on pure water and water electrolyte solutions [3]. In the first instance, conducting properties of the $H_3O^+$ and $OH^-$ ions are identified as a separated item from the cross comparison of ion separation parameters in electrolyte solutions with various chemical composition. The concentration of the $H_3O^+$ and $OH^-$ ions is further calculated under fundamental assumption that the ions do not interact.

In the above scheme, the nascent $H_3O^+$ and $OH^-$ ions (autoionization in water) are due to dissociation and association of $H_2O$ molecules. Therefore, the autoionization constant $K_w$ is equivalent to the dissociation constant.

Recently, we considered an electrodynamic response of water in the expanded frequency range to find that the spectral plateau $\sigma_\infty$ at $10^{11}$ Hz reflects the net dynamics of protons rather than the dc conductivity $\sigma_0$ (i.e. the conductivity at frequencies lower than $10^{-7}$ Hz) [16]. In this case, the high-frequency conductivity $\sigma_\infty$ (rather than $\sigma_0$, as it is accepted) is that represents the actual concentration of $H_3O^+$ and $OH^-$ ions and its temperature variation. In the present work we consider the impact of the substitution of $\sigma_0$ by $\sigma_\infty$ on the $K_w$ (pH) concept.

## 2. Experimental data

The experimental methods of obtaining $K_w$ are based on the assumption that the reversible reaction $2H_2O \leftrightarrow H_3O^+ + OH^-$ continuously goes in water. The proton dc conductivity of pure water and water electrolyte solutions is used to calculate the $K_w$ in the conductometry method. The room-temperature value of $\sigma_0$, $5.5 \times 10^{-8}$ Ohm$^{-1}$cm$^{-1}$, has been established for the dc conductivity of pure water on the bases of enormous conductometric measurements. The accepted scheme for determining $K_w$ is as follows [17].

1. Equivalent conductivity of water:
$\Lambda_w = \sigma_0 / C = 5.5 \times 10^{-8} / (55.5 \times 10^{-3}) = 0.990 \times 10^{-6}$ S cm$^2$mol$^{-1}$, where C=55.5 mol/l is a quantity of $H_2O$ molecules "dissolved" in 1 liter of water.

2. Limiting equivalent conductivity of water solution of $H_3O^+$ and $OH^-$ ions:
$\Lambda_w^0 = \Lambda^0(H_3O^+) + \Lambda^0(OH^-) = 349.8 + 198.5 = 548.30$ S cm$^2$mol$^{-1}$ (this item will be clarified below).

3. Degree of dissociation of an $H_2O$ molecule:
$\alpha = \Lambda_w / \Lambda_w^0 = 0.99 \times 10^{-6} / 548.3 = 1.8 \times 10^{-9}$.

4. Concentration of the $H_3O^+$ and $OH^-$ ions dissolved in water:
$c = \alpha \times C = 1.0 \times 10^{-7}$ mol/l

5. Ionic product (equilibrium constant, dissociation constant, autoionzation constant):
$K_w = c^2 = 1.0 \times 10^{-14}$ and, correspondently, $pK_w = -\lg K_w = 14$ and $pH = -\lg c = 7$.

Item 2 is a special one, because the determination of $\Lambda_w^0$ involves auxiliary manipulations with water electrolyte solutions. For instance, NaOH and HCl water solutions reveal, at infinite dilution, $\Lambda^0 = 426$ and 248 S cm$^2$mol$^{-1}$, respectively. Under the assumption that the electrolytes split in water with the advent of $Na^+$ and $Cl^-$ ions, the values of transport numbers, mobilities and contributions of each ion in water conductivity are determined in separate experiments. After deduction of the contributions from $\Lambda^0$, the values 349 and 199 S cm$^2$mol$^{-1}$ remain to be limiting equivalent conductivities for the $H_3O^+$ and $OH^-$ ions. Summation of this values gives $\Lambda_w^0 = 548$ S cm$^2$mol$^{-1}$, according to the Kohlrausch law by which the mobilities are summed additively in the extremely diluted solutions. This value is ascribed to water in a fundamental assumption that there is no interaction between $H_3O^+$ and $OH^-$ ions.

In our analysis, we are based on the experimental data taken from refs. [18-20]. Temperature dependences of $\sigma_0$, $\Lambda_w^0$ and $pK_w$ are presented in Fig. 1. In further discussion, these curves will be compared with other ones, which are also shown in Fig. 1. They are the temperature dependences of high-frequency water conductivity, $\sigma_\infty$, and limiting equivalent water conductivity in the form of $\Lambda'^0_w$. The latter is $\Lambda_w^0$, adjusted according to the model proposed by us in [21]. It is obtained by direct summation of limiting conductivities of binary acids and bases without taking into consideration the contributions of electrolyte ions. In our example: $\Lambda'^0_w = 426 + 248 = 674$ S cm$^2$mol$^{-1}$ instead $\Lambda_w^0 = 349 + 199 = 548$ S cm$^2$mol$^{-1}$. Thus, $\Lambda'^0_w$ is magnified, in comparison with $\Lambda_w^0$, by 10-30% with increase in temperature from 0 to 100 °C.

## 3. Analysis and discussion

The ratio of values $\sigma_\infty$ and $\Lambda'^0_w$ from items 1 and 2 gives at once (without use of 3 and 4 items) the value $c$ with the dimension mol/l which coincides with the concentration $c$ in item 5. It reflects the fact that the postulate about independence of $H_3O^+$ and $OH^-$ ions is embedded into the scheme 1-5. In the $pK_w$ (pH) concept, a simple connection of the conductivity $\sigma_0$ with the concentration $c$ of the $H_3O^+$ and $OH^-$ ions and with their mobility, $\mu = \Lambda^0_w/q$, is accepted: $\sigma_0 = cq\mu$, where $q$ is the elementary charge (the $H_3O^+$ and $OH^-$ ions mobilities are not distinguished for better clarity of the scheme). A payment for the simplicity is a challenge to explain the temperature behavior of the concentration $c$.

As seen from Fig. 1, the temperature dependence of $pK_w$ results from different temperature runs of the curve for conductivity, $\sigma_0(T)$, and the curve for mobility, $\Lambda^0_w(T)$. The values $\sigma_0$ and $\Lambda^0_w$, depending on the temperature, behave as if they would reflect different microscopic mechanisms of the conductivity. Indeed, though the $H_3O^+$ and $OH^-$ ions are the only charge carriers in an water electrolyte solution at its infinite dilution, their activation in conductivity (temperature change) keeps specifics of an electrolyte: activation energy of water electrolyte solutions conductivity is maintained at any concentration, including the case of infinite dilution [20]. It follows that the temperature dependence of $K_w(T)$, along with that it reflects a tendency of an $H_2O$ molecule to autodissociation, contains imprint of the temperature behavior of the electrolyte.

An opportunity to untie the properties of water and electrolyte is seen in considering of the temporary characteristics of the dissociation-association processes. In ref. [16] we identified, in the spectrum of dynamic conductivity of water, the two dispersion-free plateaus, strongly different-scale on frequency: $\sigma_0$ (at lower than $10^7$ Hz) and $\sigma_\infty$ (at higher than $10^{11}$ Hz, see Fig. 1 in [16]). We interpreted such a shape of the spectrum as manifestation of high concentration of the counter ions being in the atmosphere of strong interactions. With such an approach, the high-frequency shelf $\sigma_\infty$ is a response of charges free from interaction ("bare charges"), while the dc conductivity $\sigma_0$ is a response of the same, but "dressed" charges. True concentration of the $H_3O^+$ and $OH^-$ ions is represented by the conductivity of the bare charges, $\sigma_\infty$.

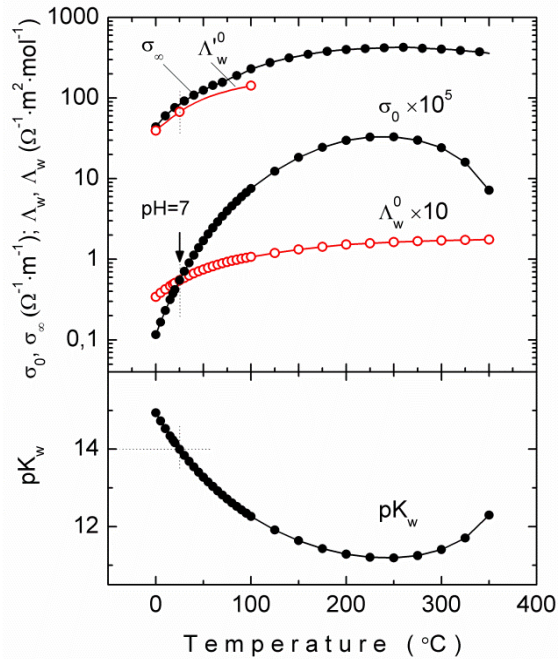

**Figure 1.** Temperature dependencies of the electrical parameters of water: the specific conductivities $\sigma_0$ [18] and $\sigma_\infty$ [19]; the equivalent conductivities $\Lambda^0_w$ [20] and $\Lambda'^0_w$ (see text); the coefficient of the autoionization constant $pK_w$ [18]. The dotted lines mark T = 25 °C, the temperature at which $pK_w=14$ and pH=7.

Additionally, we have found that in water electrolyte solutions the position of the $\sigma_0$ plateau changes by its pulling up to the $\sigma_\infty$ shelf during increase of the electrolyte concentration (Fig. 1 in [21]). As to the

$\sigma_\infty$ shelf, its position does not change with change in concentration, always remaining near the $\sigma_\infty$ shelf of pure water. We associated the $\sigma_0$ pulling up with activation in the dc conductivity of water $H_3O^+$ and $OH^-$ ions. They inherently exist in water but are "dressed" in the interaction atmosphere. The electrolyte "undresses" such $H_3O^+$ and $OH^-$ ions, making them visible in the dc conductivity $\sigma_0$.

Taking into account the aforesaid, let us choose $\sigma_\infty$ instead of the common $\sigma_0$ as a measure of the concentration of $H_3O^+$ and $OH^-$ ions in water. Also, let us use $\Lambda'^0_w$ instead of $\Lambda^0_w$. Thus doing, we pass from consideration of the lower pair of the curves in Fig. 1 ($\sigma_0(T)$ and $\Lambda_w^\infty(T)$) to the top pair ($\sigma_\infty(T)$ and $\Lambda'^0_w(T)$).

It must be particularly emphasized that the $\Lambda'^0_w$ value, which is, though, an attribute of pure water, is artificially constructed from the data on dc conductivity of water electrolyte solutions. By our interpretation given in ref. [21], the role of the electrolyte in a water solution is activating of water ions, $H_3O^+$ and $OH^-$, rather than delivering of its own ions into the solution. Equivalence of the activation energy $\Lambda'^0_w(T)$ and the activation energy of the high-frequency conductivity $\sigma_\infty(T)$, as shown in Fig. 1, confirms the assumption that exactly $H_3O^+$ and $OH^-$ ions shall be identified as conducting ions.

As is seen, $\sigma_\infty$ and $\Lambda'^0_w$ change, depending on the temperature, in a similar way. Approximately the same slop of the curves $\sigma_\infty(T)$ и $\Lambda'^0_w(T)$ reveals a very slight change in their correlation with the temperature change. It is of fundamental importance that in a control temperature point 25 °C, the ratio of values $\sigma_\infty = 79$ S m$^{-1}$ and $\Lambda'^0_w = 674$ S cm$^2$mol$^{-1}$ gives, for the concentration of $H_3O^+$ and $OH^-$ ions in water, the value $c=79/0.0674=1.17$ mol/l which is close to the one which was found earlier by us in ref. [16] (~ 1 mol/l). Thus, it follows from two different approaches, without use and with use of electrolytes, that water looks as if ~ 1% of water molecules would be separated into the $H_3O^+$ and $OH^-$ ions.

Formally, one can introduce, by analogy with the common $K_w=c^2$, an ionic product $K'_w=c_\infty^2$ where $c_\infty(T) = \sigma_\infty(T)/\Lambda'^0_w(T)$. Then, within the range of 0-100 C, the $K'_w$ changes from 1.2 to 1.6.

However, $K'_w$ relates to intactness of a water molecule specifically. This parameter gains a physical sense, if one speculates that the $H_3O^+$ and $OH^-$ ions, being in high concentration, create for each other the dynamical interaction potential. Neutral $H_2O$ molecules are exposed by this potential also [16]. In accordance to NMR studies and MD simulations, the ions and neutral water molecules are in permanent proton exchange on the picosecond time scale [6, 7, 9-12]. And if, as found by us, the concentration of $H_3O^+$ and $OH^-$ ions dos not depend on temperature, exactly this process, rather than autodissociation, produces the electrodynamic spectral response. Within this scheme, $K'_w$ characterizes the stable concentration of $H_3O^+$ and $OH^-$ ions being in the dynamic interaction potential, while the common $K_w$ constant, represents, figuratively, an ionic vapor density over this potential. The concentration of this ionic vapor is that detected in common conductometric measurements.

The concentrations of molecules and ions are in thermodynamic equilibrium in water. Given that the lifetime of ions is measured by picoseconds [11, 12] and their concentration is of 1%, as is found by us, the lifetime of an $H_2O$ molecule is ~ 0.1 nanosecond (instead of the accepted ~10 hours [6, 7]).

## 4. Conclusion

We have considered the dissociation-association processes in water taking into account their time-scales. Namely, we proceeded from the assumption that a net proton dynamics in water is represented by the high-frequency plateau $\sigma_\infty$ (higher than $10^{11}$ Hz), rather than the low-frequency plateau $\sigma_0$ (lower than $10^7$ Hz) within the spectrum of water conductivity, as commonly assumed. We confirmed by independent way than earlier that the concentration of the counter $H_3O^+$ and $OH^-$ ions in water is of 7 orders higher than that commonly accepted for pH=7. We found this concentration to be not dependent on temperature to conclude that water electrodynamics it is largely due to a proton exchange between $H_3O^+$ (or $OH^-$) ions and neutral $H_2O$ molecules rather than spontaneous ionization of $H_2O$ molecules. In our interpretation, the common $K_w$ constant (or pH) characterizes the thermoactivation of $H_3O^+$ and $OH^-$ ions from the potential of their mutual interaction.

## 5. Acknowledgments

The authors thank A. V. Pronin for close collaboration on related topics.

## 6. References

[1]   Eisenberg D. and Kauzman W., *The Structure and Properties of Water* (Clarendon Press, Oxford, UK) 1969.
[2]   Franks F., *Water: A Comprehensive Treatise* (Plenum Press, New York) 1972–1982.
[3]   Glasstone S., *An Introduction to Electrochemistry*, fourth edition (East West Press PVT. LTD, New Delhi, India) 1974.
[4]   Bockris J. O'M. and Reddy A. K. N., *Modern Electrochemistry 1: Ionics*, 2nd edition (Kluwer Academic Publishers, New York) 1998.
[5]   Bates R. G., *Determination of pH. Theory and Practice*. John Wiley & Sons Inc. NY, London, Sydney, 1964.
[6]   Hassanali A., Prakash M. K., Eshet H. and Parrinello M., *Proc. Natl. Acad. Sci. U.S.A.,* 108 (2011) 20410.
[7]   Bakker H. J. and H.-K. Nienhuys *Science* 297 (2002) 587.
[8]   Eigen M. and L. De Maeyer, *Z. Elektrochem.* 59, (1955) 986.
[9]   Agmon N., *Account of Chemical Research* **45**, N1 (2012) 63.
[10]  Fernandez-Serra M. V. and Artacho E., *Phys. Rev. Lett.*, **96** (2006) 016404.
[11]  Marx D., Chandra A., Tuckerman M. E., *Chem. Rev.* 110 (2010) 2174.
[12]  Walbran S. and A. A. Kornyshev *J. Chem. Physics* **114** (2001) 10039.
[13]  Cabane B. and Vuilleumier R. *The physics of liquid water.* Elsevier, 2005, 337, pp.159.
[14]  Bandura A. V. and Lvov S. N., *J. Phys. Chem. Ref. Data, Vol.* 35, No. 1, 2006 15-30.
[15]  Yagasaki T., Iwahashi K., Saito S., Ohmine I., *J. Chemical Physics* **122**, 144504 (2005).
[16]  Volkov A. A., Artemov V. G., Pronin A. V., *EPL,* **106** (2014) 46004.
[17]  Rajeev J., *Physical Chemistry, Electrochemistry I*, School of Studies in Chemistry, Jiwaji University, Gwalior – 11 www.academia.edu 2015.
[18]  Light T. S., Licht S. L., Anal. Chem. 59 (1987) 2327.
[19]  Scherbakov V., Artemkina Y., Ermakov V., *Electrolyte solutions.* Palmarium Acad. Publishing, Saarbrucken (2012) 132 (in Russian).
[20]  *CRC Handbook of Chemistry, and Physics,* 70th Edition, Weast, R. C., Ed. (CRC Press, Boca Raton, FL) 1989, D-221.
[21]  Artemov V. G., Volkov A. A., Sysoev N. N., Volkov A. A., *EPL*, **109** (2015) 26002.
[22]  Meiboom S., *J. Chem. Phys.* **34** (1961) 375.